\documentclass[preprint2]{aastex62}
\graphicspath{{./}{figures/}}
\pdfoutput=1
\received{October 18, 2018}
\shorttitle{On the orbit distribution of distant TNOs}
\shortauthors{Kavelaars et al.}

\begin{document}

\title{Perspectives on the distribution of orbits of distant Trans-Neptunian Objects}

\correspondingauthor{J. J. Kavelaars}
\email{JJKavelaars@gmail.com}
\author[0000-0001-7032-5255]{J. J. Kavelaars}
\affil{Department of Physics and Astronomy, University of Victoria, Elliott Building, 3800 Finnerty Rd, Victoria, BC V8P 5C2, Canada}
\affil{Herzberg Astronomy and Astrophysics Research Centre, National Research Council of Canada, 5071 West Saanich Rd, Victoria, British Columbia V9E 2E7, Canada}

\author[0000-0001-5368-386X]{Samantha M. Lawler}
\affil{Herzberg Astronomy and Astrophysics Research Centre, National Research Council of Canada, 5071 West Saanich Rd, Victoria, British Columbia V9E 2E7, Canada}

\author[0000-0003-3257-4490]{Michele T. Bannister}
\affiliation{Astrophysics Research Centre, School of Mathematics and Physics, Queen's University Belfast, Belfast BT7 1NN, United Kingdom}

\author[0000-0002-3507-5964]{Cory Shankman}
\affiliation{Department of Physics and Astronomy, University of Victoria, Elliott Building, 3800 Finnerty Rd, Victoria, BC V8P 5C2, Canada}

\begin{abstract}

Looking at the orbits of small bodies with large semimajor axes, we are compelled to see
patterns. 
Some of these patterns are noted as strong indicators of new or hidden processes in the outer Solar System, others are substantially generated by observational biases, and still others may be completely overlooked. 
We can gain insight into the current and past structure of the outer Solar System through a careful examination of these orbit patterns.
In this chapter, we discuss the implications of the observed orbital distribution of distant trans-Neptunian objects (TNOs).  
We start with some cautions on how observational biases must affect the known set of TNO orbits. 
Some of these biases are intrinsic to the process of discovering TNOs, while others can be reduced or eliminated through careful observational survey design.
We discuss some orbital element correlations that have received considerable attention in the recent literature. 
We examine the known TNOs in the context of the gravitational processes that the known Solar System induces in orbital distributions. 
We discuss proposed new elements of the outer Solar System, posited ancient processes, and the types of TNO orbital element distributions that they predict to exist.  
We conclude with speculation.

\end{abstract}


\section{Biases in the detection of distant Solar System objects}
\label{sec:intro}

The Kuiper belt is over 4.5 billion km distant from the Earth-bound observer, with the most distant trans-Neptunian objects (TNOs) known being three times further away still.  The challenge of detecting objects at these great distances should not be underestimated.  The Sun's light reflected of Solar System bodies at a distance $r$ is dimmed by the factor $r^{-4}$, greatly exaggerating our sensitivity to nearby objects in comparison to more distant objects.  The volume of the Solar neighbourhood that a survey is sensitive to, its {\it detection volume},  is, at minimum, limited in radial extent. The strength of the $r^{-4}$ observational bias is frequently under appreciated when attempting to interpret the distributions of objects detected by a particular survey.

TNOs on orbits with moderate to large eccentricities also present a distorted view of the population.  Eccentric TNOs occupy a range of Solar distances during their orbits, resulting in a time-variable $r^{-4}$ flux bias. A TNO may only spend a small fraction of its orbital period within the detection volume of a particular survey. The larger the semi-major axis, the larger the eccentricity needed to bring the object within the detection volume, and the smaller the fraction of the orbit for which that object remains in the detection volume.

Limited telescopic resources add another layer of complexity to the problem of quantifying the biases inherent in the detected sample of TNOs. One can only detect objects in the part of the sky where one looks.  This observer direction bias imposes a relation between the nodal angle ($\Omega$), argument of pericentre ($\omega$), mean anomaly ($\mathcal{M}$) and inclination ($i$) of the orbits that can be detected. This coupling of multiple angles can be difficult to conceptualize.  For illustration, consider a discovery survey whose fields all straddle the ecliptic plane.  In those surveys,  orbits that are inclined to the ecliptic plane with an inclination that exceeds the latitude sensitivity of the survey will only be detectable when looking towards their nodes. Combine the forced detection at the node with the preferential discovery of objects near their pericentre $q$, caused by the $r^{-4}$ flux bias discussed above, and such a survey will find that the detected sample of TNOs on inclined orbits all have arguments of pericentre $\omega$ near 0$^{\circ}$ and 180$^{\circ}$. This effect is well known, and is described here to remind the reader of the basic processes at work.  When attempting to maximize the science return of scarce observing time, observing fields in specific areas of sky inherently induces biases in the detected sample, and some of these biases may be difficult to recognize. 

\begin{figure*}
\includegraphics[width=\textwidth]{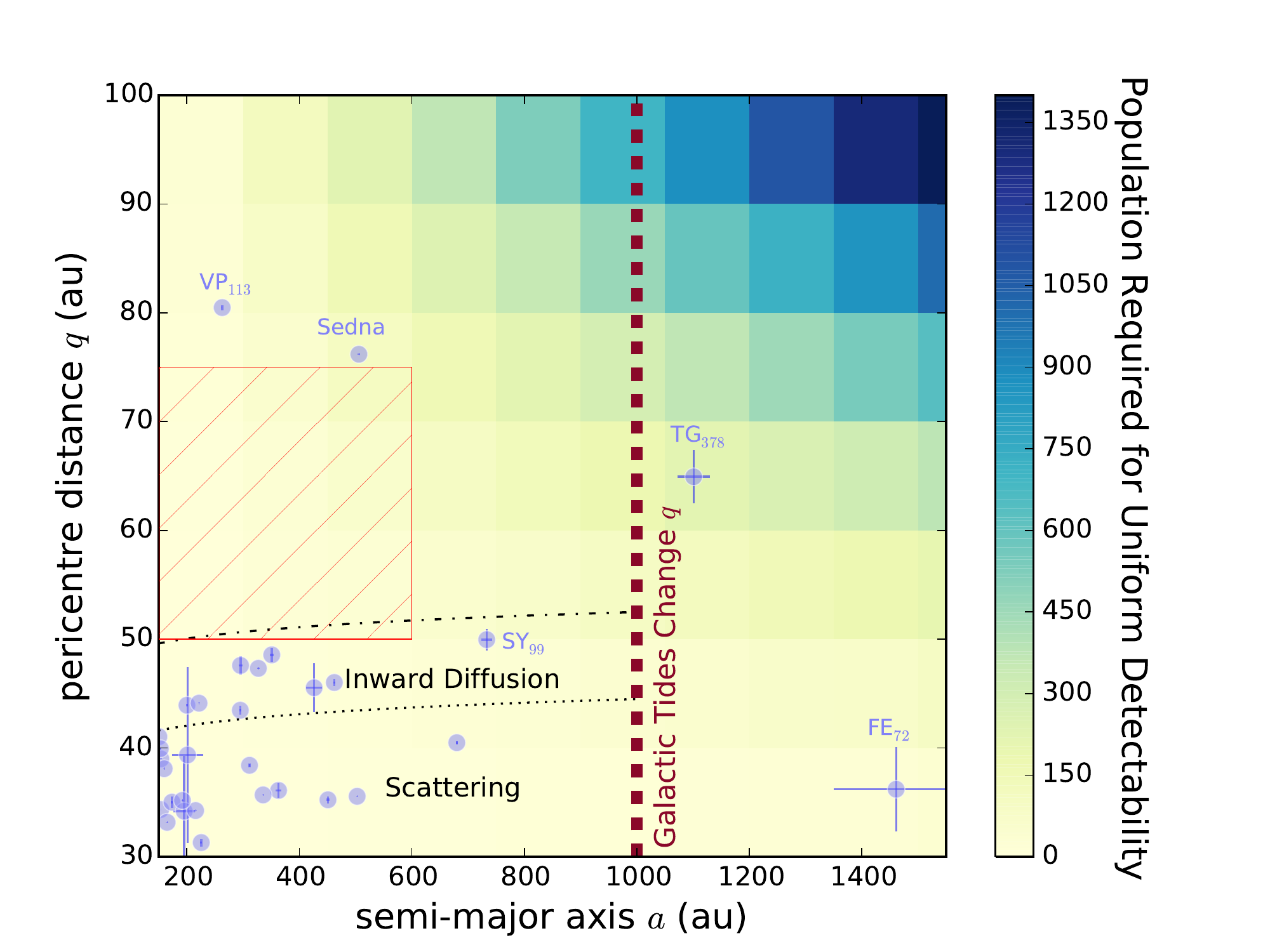}
\caption{Orbital distribution of known TNOs with observed arcs longer than 10 months in the MPC Database as of October 2018, on orbits with $a > 150$~au and $q > 30$~au in $a$-$q$ (blue dots); a few extremes are noted by name, uncertainty in orbital parameters extracted from orbfit \citep{bernstein00}. 
The background grid of coloured boxes indicates the population of TNOs needed in each ($a$, $q$) bin for detection of that orbit to have similar probability if there is a single object in the $a = 150$, $q = 30$ bin; see text for details.
Horizontal curves approximate the boundary below which inward diffusion from $a > 1000$~au is significant \citep[dot-dash line;][]{bannister17}, and the boundary below which outward diffusion (chaotic scattering) in $a$ is significant (dotted line). 
The thick dashed vertical line roughly indicates where Galactic tides become a notable long-term influence on the percentres of orbits.
The red hatched box indicates a region of ($a$, $q$) orbits that currently has no TNO detections, though detectability is actually easier here than for the two known highest-$q$ TNOs, Sedna and VP$_{113}$.
}
\label{fig:bias}
\end{figure*}

In Figure~\ref{fig:bias} we present the orbital distribution of a subset of the trans-Neptunian objects reported to the Minor Planet Center (MPC) as of 2018-Oct-15: all TNOs with $a > 150$~au and $q > 30$~au.  The vertical line at 1000~au {\em roughly} indicates the more distant phase space where the effects of Galactic tides and stellar passages become important \citep[e.g.][]{kaib09}. The horizontal dotted curve indicates the zone below which outward diffusion (chaotic scattering) from the Kuiper belt is significant. Between that and the dot-dashed curve indicates the zone where inward diffusion from the inner Oort cloud occurs (see Section~\ref{sec:diffusion} and \citealt{bannister17} for details).  Also shown in this figure is background colour-coding estimating the size of the intrinsic population that would be needed to detect an single object on a given ($a, q$) orbit, in a survey that also detected one object with $a \sim 150$~au and $q \sim 30$~au (assuming a size-frequency distribution of TNOs $ \Sigma{N} = 10^{0.5(H-H_o)}$).  From this figure, we can see that the present low detection rate in the $q>60$~au, $a > 1000$~au orbits is only a weak constraint on the size of that population as our ability to see into this zone is quite limited. In order to detect one TNO in this ($a$,$q$) range, a survey that detected one TNO at low-$a$ and low-$q$ would require a couple hundred times as many TNOs with obits in the large ($a$, $q$) zone. 
We would need hundreds of TNO detections in the low-($a$,$q$) zone just to rule out a uniform distribution in this phase-space.  
These numbers are in basic agreement with more careful computation provided elsewhere \citep[e.g.][]{sheppard18} and are given here to guide the reader's understanding of the influence of orbit and flux bias in the detected sample.
An important consideration is that to use the biases in Figure~1 to aid in understanding the structure of the trans-Neptunian region, we need to know the full range of orbits, including those with $a< 200$~au and $q<40$~au, that were detected in a given survey. We can then use the relative sensitivity to scale between the regions. 

Given the sample of TNOs that are publicly known, one is tempted to pursue mechanisms to debias the observed sample.  Two classes of approaches are common.  In the first, the characteristics of the survey itself are used to determine the efficiency of detections of various orbits, this approach has been employed in numerous project \citep[e.g.][]{trujillo00,elliot05,petit11,adams14,bannister18}.  A  weakness of this approach is that each project has taken some what different approaches to documenting their characterization and this makes combining datasets, to enhance statistical power, difficult.   Recently \citet{brown17}, and continuing in \citet{brown19}, have implemented procedures that attempt to  `self characterize' the detection biases present in the Minor Planet Centre reported sample of KBOs. This approach has the advantage that it combines together a larger dataset but has some disadvantages also. 

Using the public catalog to attempt to debias the detected sample makes at least two implicit assumptions that are not true: that the sky coverage of a survey is well represented by the detections in that survey and that all objects detected by a survey are report.

Using the public catalog provides no capacity to know where surveys did not detect objects.  As an example, consider the Canada France Ecliptic Plane Survey High Latitude Component \citep{petit17}.  CFEPS-HiLat imaged 700 square degrees of sky searching for TNOs on high-inclination orbits but only reported the detection of 24 objects.  This survey provides constraints on the number of objects that can be on highly inclined orbits, a constraint that is not visible if one uses the known objects as probes of the locations of sky that have been surveyed.

An additional weakness is that the use of the detected sample to determine the characterization assumes that there is no reporting bias (ie. all objects detected are tracked and reported), but this is known to not be the case.  Some projects are forced, by the nature of resource restrictions, to only report and track a selected sample of their detections (such as only reporting the detection of objects beyond some defined distance from the observer,  \citep[e.g][]{sheppard18} ), such reporting bias is impossible to determine from the detected sample and can lead to significant misinterpretation. 

The desire for a statistically useful sample can now, largely, be achieved by using the sample provided by the `OSSOS Ensemble' \citep{bannister18} which includes orbits and detection circumstances for 1086 Kuiper belt objects, about 40\% of the currently known population of KBOs observed at 2 or more oppositions.  

One point that is worth noting is the lack of detections in the region shown by the red hatched box in Figure~1.  The hatched box region is devoid of known TNOs --- however, our sensitivity to orbits in this zone is similar to that in other zones, where in contrast a number of detections exist. 
This may indicate that this zone is, indeed, relatively under-populated, an important point in constraining the dynamics of this region of the Solar System.  Or this may be a region of phase space where surveys searching for distant TNOs have simply culled  their detections in an effort to concentrate resources on tracking the large peri-centres, distant, members. The apparent paucity of orbits in this zone has been noted previously \citep[e.g.][]{trujillosheppard14,bannister18}, and is discussed further in Section~\ref{sec:highq}.

\section{Potential mechanisms forming the orbits of high-pericentre TNOs} \label{sec:highq}
In recent literature, there has been much discussion of the handful of known large-$a$, high-$q$ so-called ``extreme TNOs,'' which are defined with a variety of orbital selection criteria\footnote{We note that definitions that make use of orbital elements for such orbits must use the barycentric not the heliocentric orbit, due to the long-term effect of Jupiter on these distant orbits.
However, heliocentric orbital elements are what is provided by the most frequently used databases of orbits, the Minor Planet Center Database and JPL Horizons. Uncertainties in the orbit fit to the measured TNO astrometry should also be considered.},  sometimes as having orbits with $a>250$~au (alternatively sometimes $a>150$~au) and perihelia detached from active interaction with Neptune of $q\gtrsim37$~au \citep[e.g.][]{Kiss13,Sheppardetal2016,dlfmdlfm16c,shankman17bias,bannister17,becker18}.
This definition is vague, reflecting the current lack of understanding of the population, and may not necessarily be dynamically distinguishing. 
\citet{batygin19} provide a lengthy review of the `Planet 9' hypothesis and provide some dynamical considerations on where the boundary should be drawn when considering dynamics that could be induced by a large external planet.  
They draw the boundaries at $q > 30$~au and $a > 250$~au.
As of October 2018, 17 TNOs with multi-opposition orbits reported in the Minor Planet Center database have $150<a<1000$~au with $q>37$~au, mostly with pericentres in the range $37<q<50$~au.
Two of these TNOs have much larger pericentre distances, with $q>75$~au: Sedna ($q=76.19 \pm 0.03$~au, $a=507 \pm 10$~au; \citealt{brownetal04}) and 2012~VP$_{113}$ ($q=80.3^{+1.2}_{-1.6}$~au, $a = 266^{+26}_{-17}$~au; \citealt{trujillosheppard14})\footnote{Barycentric orbits after \citet{bannister17}.}. 

It has been long recognized that the orbital distribution of TNOs can be used to understand the past dynamical history of the Solar System, particularly the outer giant planets \citep[e.g.][]{malhotra93,levison08}.  
See \citet{nesvorny18} for a recent review of this topic.
The basic origin scenario is that TNOs formed in a dynamically cold disk of planetesimals, which was largely disrupted when Neptune migrated, placing TNOs into classes based on orbits that display unique dynamical behaviours \citep[see][]{gladman08}.  
High-$q$ TNOs with $a<1000$~au are difficult to explain in this framework: they never approach Neptune closely enough to receive the strong dynamical kicks needed to change their orbits, their large eccentricities preclude in-situ formation, and their orbits are not large enough to be affected significantly by Galactic tides.  

In a sign of a vigorously active area of theoretical investigation, the high-pericentre TNOs have recently spawned a flurry of studies to explain these dynamically interesting orbits.
While attention has focused on the hypothesis that an undiscovered distant giant planet can be used to explain properties of large-$a$, high-$q$ orbits \citep[see][for a recent review]{batygin19}.  While the Planet 9 hypothesis provides a compelling explanation for various orbital characteristics, a planet presently orbiting in the distant Solar System is by no means the only theory being advanced.
We highlight several proposed classes of theories that can raise TNO orbital pericentres.

Simulations that include perturbations by passing stars at various times in Solar System history produce high-pericentre TNOs on large-$a$ orbits.
Stellar perturbations have been considered to raise perihelia of native TNOs while the Sun is still in the denser stellar environment of its birth cluster with lower relative velocities \citep{Fernandez:2000, kenyonbromley04, morbidellilevison04,  Brasser:2006, KaibQuinn2008, brasser12, brasserschwamb15, Pfalzneretal2018}, by field stars during the Sun's post-cluster orbits of the Galaxy, and possible radial migration \citep{kaib11b}.
Certain geometries of stellar flyby may permit capture of TNOs from passing star systems (\citealt{kenyonbromley04,morbidellilevison04,Jilkovaetal2015}; see also \citealt{Levison:2010} for Oort comet capture).
All these simulations have many degrees of freedom related to the stellar mass, and the distance and geometry of a flyby, which require an abundance of known high-$q$ TNOs to well confine their possible parameter space.

An intriguing new theoretical mechanism is being explored by simulations that take into account the self-gravitational influence of great densities of small TNOs. \citet{madigan2016} have shown an ``inclination instability'' within a massive planetesimal disk can produce high-$q$ orbits. A disk that begins as axisymmetric eccentric orbits like that of the scattering disk will increase in inclination while lowering its orbital eccentricities, forming an asymmetric cone \citep{Madigan:2018}.
The inclination instability effect requires substantial mass in planetesimals in the distant Solar System for it to initiate; \citet{Madigan:2018} infer about half an Earth
mass at hundreds of au.  
\citet{sefilian19} have further explored this mechanism and find that a massive disk would induce clustering in orbital angles and argue that massive disk may be more likely than expected as such disks are common around other stars.
 \citet{batygin19}, however, expect that such a massive disk unlikely to have remained in place for the age of Solar System, making this explanation of clustering unlikely to be correct. 
Bounds on the size distribution of distant TNOs and thus the mass at large $a$ remain limited, but will constrain this theory tightly in the future. 

Two mechanisms can produce many high-$q$ TNOs solely from the known planets of the Solar System.
First, chaotic diffusion in semimajor axis caused by weak gravitational kicks from Neptune can cause minor planets to migrate from the inner Oort cloud to large-$a$, high-$q$ TNO orbits \citep{Duncan1987, Kaib:2009, bannister17}; we discuss this further in Section~\ref{sec:diffusion}.
Second, high-$q$ orbits may be produced during Neptune's migration. 
TNOs that are captured into Neptune's mean-motion resonances (MMR) experience Kozai oscillations inside the MMR. 
As Neptune migrates outwards, they may drop out of the resonance at high-$q$, where the resonance is narrower. 
The TNO is then ``fossilized'' on a dynamically detached, long-term stable orbit \citep{gomes03}.
Several non-resonant, stable TNOs have been discovered on high-$q$ orbits near strong resonances, lending observational support to this theory \citep{pike15,lawler18res}.
Including dwarf planets in migration simulations (cf.\ ``grainy" migration; \citealt{nesvorny16}), as required from the size distribution of the initial disk, is a recent refinement that is still being explored, but appears important.
Grainy simulations show that the mode and timescale of Neptune's migration affects the distribution of these high-$q$ resonant dropouts \citep{Nesvornyetal2016,kaib16}.
While simulations of scattering TNO capture into MMRs show that this can be effective for raising pericentres as high as $q\gtrsim70$~au, this can only happen for scattering TNOs that already have large inclinations prior to resonant capture \citep{gallardo12}, and while grainy migration can modify the inclination distribution \citep{marco18} it may not be sufficient and so this may not be the emplacement mechanism for Sedna and VP$_{113}$ at $i<25^{\circ}$, despite their locations near low-order, distant resonances.  

\begin{figure*}

\includegraphics[width=\textwidth]{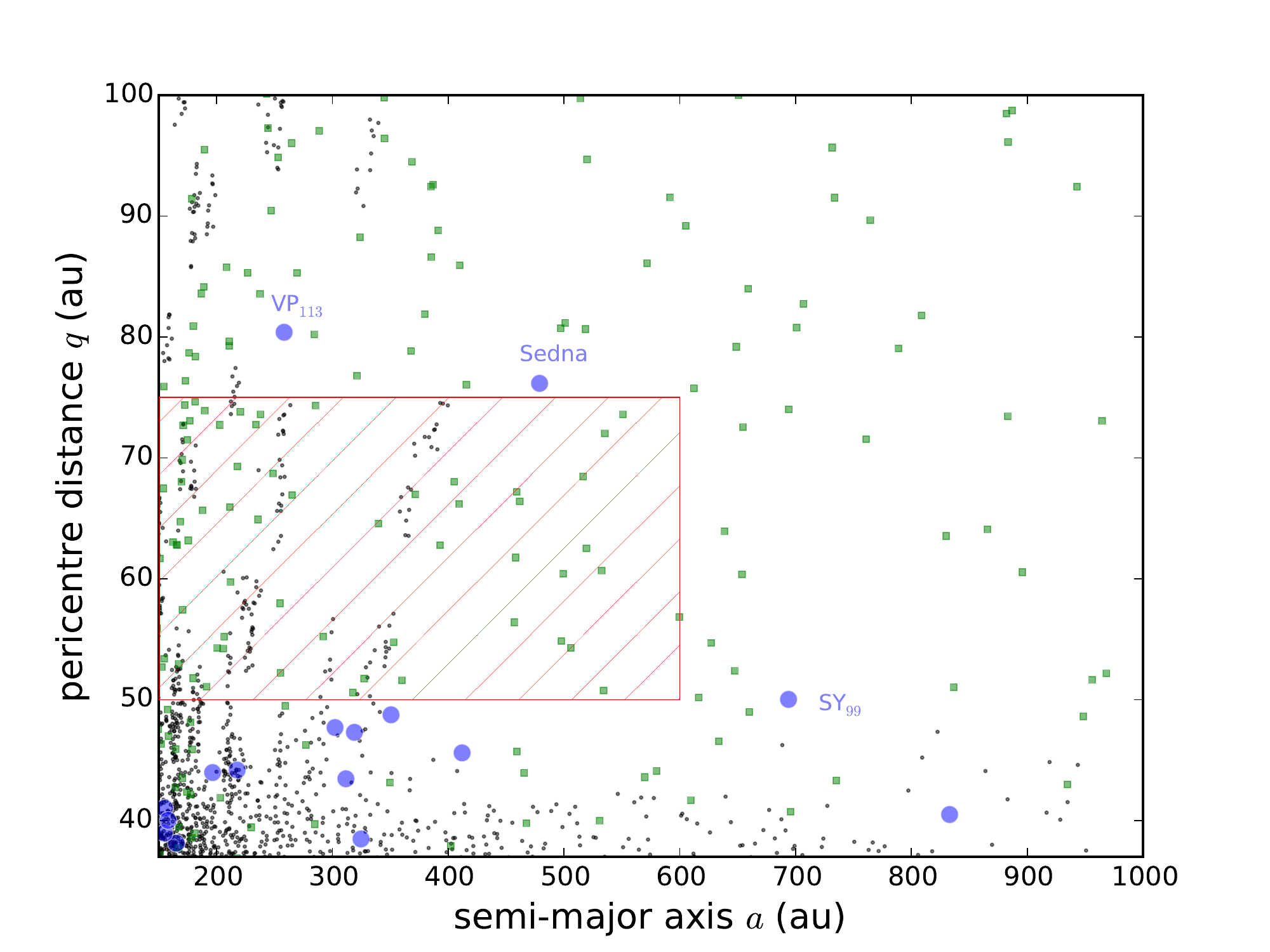}
\caption{Orbital distribution of simulated TNOs resulting from the rogue planet simulation in \citet{gladmanchan06}, cloned and integrated for an additional Gyr (small black points), and simulated TNOs resulting from an emplacement simulation of \citet{lawler2017} including an eccentric distant giant planet (green squares).
As in Figure~\ref{fig:bias}, large blue dots show real TNOs and the red hatched box indicates the region of ($a$, $q$) orbits yet without real TNO detections.
Both the rogue planet model and the additional planet model are able to produce TNOs with very high $q$ values, even higher than 2012~VP$_{113}$ and Sedna, and 
both models also produce TNOs inside the hatched box. 
If either of these models represents reality, Figure~\ref{fig:bias} shows that it is very unlikely that there would be zero detections of TNOs inside the box, due to the greater detectability of TNOs in that box as compared with 2012~VP$_{113}$ and Sedna.
}
\label{fig:rogue}
\end{figure*}

A separate class of hypotheses invokes planetary-mass bodies to raise pericentres.
The possible presence of an undiscovered massive distant planet has been discussed extensively in the literature from the early days of Kuiper Belt discoveries to the present \citep{gladman02,brownetal04,Gladman:2005,LykawkaMukai2008,soaresgomes13,trujillosheppard14}.
Many recent simulations have shown that a distant massive planet would be quite effective at raising the pericentres of large-$a$ TNOs \citep[][see Figure~\ref{fig:rogue}]{batyginbrown16,shankman17,lawler2017,Lietal2018}.
However, other aspects of the observed Kuiper Belt are solidly inconsistent with this particular planetary scenario \citep{lawler2017,shankman17,shankman17bias,nesvorny17}. 
Simulations that include one or more ``rogue planets'', with masses similar to Mars or Earth that are ejected after orbiting in the Kuiper Belt region for a few hundred million years, are also very successful at lifting pericentres for large-$a$ TNOs \citep{gladmanchan06,silsbee18}.

Figure~\ref{fig:rogue} shows the TNO orbits resulting from a rogue planet simulation where a 2~Earth mass planet started at $a=35$~au and $q=30$~au, and was ejected by interactions with the giant planets after 200~Myr \citep[from][]{gladmanchan06}.  
The TNO test particles are then integrated for 1~Gyr in the presence of the four giant planets, resulting in the distribution shown (the clustering results from cloning of particles, prior to the 1~Gyr simulation).
Also shown for comparison in Figure~\ref{fig:rogue} is a TNO emplacement simulation from \citet{lawler2017} that includes an eccentric distant planet.
For comparison, the same real TNOs are plotted as in Figure~\ref{fig:bias}.
We note that this is provided as an indicative rather than a fully quantitative comparison, as survey characterizations do not exist for all of the discovery surveys of this ensemble of known TNOs, and thus cannot be applied to appropriately bias the simulation outcomes.
While both the rogue planet model and distant giant planet model manage to produce TNOs at very high pericentre distances, even higher $q$ than 2012~VP$_{113}$ and Sedna, both models also produce many TNOs inside the red hatched box, which contains zero real TNO detections to date.
The TNO distribution produced by either of these models predicts that there should be more easily detectable TNOs inside the red box, and it is unlikely that the highest-$q$ TNOs would be detected without any TNOs detected in the $50~{\rm au} < q < 70$~au range.
With either of these models, 2012~VP$_{113}$ and Sedna remain hard-to-explain outliers.

All of these hypotheses for lifting pericentres of distant TNOs have associated simulations modelling a small-body population, produced with many degrees of freedom. 
The critical test for each hypothesis is how well it reproduces the observed TNO population. 
Several of these models are currently providing population outcomes at the level of detail necessary for testing against the observed TNO population (cf.\ Figure~\ref{fig:rogue}); others are still maturing toward that critical point.

\section{Diffusion and motion of large semi-major axes orbits} \label{sec:diffusion}

The nuanced effects of gravitational perturbation from the planets extend over remarkably wide spatial scales and timescales for large semimajor axes orbits, in ways not seen in the inner Solar System.
As initially suggested by \citet{Duncan1987}, each distant encounter of Neptune by a TNO on a near-parabolic $a \gtrsim 100$~au orbit with a perihelion exterior to Neptune will produce an energy change in the TNO's orbit, even for high-$q$ orbits. 
The effect of the energy change at each perihelion passage is a change in the size of the orbit's semimajor axis, while the orbit's perihelion stays constant.
The weak kicks by Neptune at the TNO's perihelion change its orbital $a$ on a timescale $\propto a^{-1/2}$.
As the semimajor axis changes can be modelled as a random walk, with the orbit either becoming larger or decreasing in size with each passage, this change in orbital dimensions for large-$a$ detached TNOs is an example of dynamical diffusion.
It is an effect that occurs purely under the gravitational influence of the known planets.

\citet{bannister17} showed that diffusion is a substantive effect over Gyr for the large-$a$ detached (high-$q$) TNO orbits.
This investigation was prompted by the discovery of 2013 SY$_{99}$ in the course of the OSSOS survey, on an orbit with $q = 50.0$~au, $a = 733 \pm 42$~au (noted on Figure~\ref{fig:bias}).
Perihelia passages for orbits as large and distant as SY$_{99}$'s are only every 20~kyr, thus the energy walk is slow and requires the passage of Gyr to show changes in orbital semimajor axis.
The semimajor axis of SY$_{99}$ can change by a factor of two over the age of the Solar System, due to the semimajor axis diffusion of 100~au or more that it experiences on Gyr timescales --- despite being fully 20~au separate at perihelia passages with Neptune's orbit.
There were hints of the presence of diffusion in earlier studies of large TNO orbits: \citet{gladman02} found diffusive chaos when examining the orbit of 2001 $CR_{105}$ (the first known member of the extreme orbit group), \citet{sheppardtrujillo16} noted semimajor axis mobility in the orbit of their discovery 2013~FT$_{28}$ ($q = 43.47 \pm 0.08$~au, $a = 295 \pm 7$~au), while \citet{gallardo12} and \citet{brasserschwamb15} saw diffusion in their modelling of sub-samples of extreme TNO orbital phase space.
Integrations of the then-known $45<q<50$~au TNOs, with $180 < a < 300$~au, in the presence of the giant planets showed that they exhibit diffusive semimajor axis behaviour \citep{bannister17}. 
Like Sedna and 2012 VP$_{113}$, these orbits are within the placid $a \lesssim 1000$~au region where they are isolated from the Gyr-timescale influence of perturbations by the Galactic tide \citep[e.g][]{brasserschwamb15}.

The evaluation of the largest minor planet orbits cannot take place in isolation. 
Orbits at several thousand au start to experience the effects of the Galactic tide, in the inner fringe of the Oort cloud \citep{dones04}.
The population density of the $a \sim 2000$~au inner Oort cloud region is presently largely mysterious. 
Long-period comets are sourced from several tens of thousands of au, where the influence of the Galactic tide is dominant \cite[e.g.][]{dones04, vokrouhlick19}.
The apocentres of scattering disk member orbits can extend into the inner Oort region, such as that of 2014~FE$_{72}$ ($q = 36.3\pm 0.1$~au, $a=1505\pm 540$~au\footnote{heliocentric JPL Horizons elements from a 1511 day arc, computed 2018-Jun-11.}; noted in Figure~\ref{fig:bias}; \citealt{sheppardtrujillo16}). While other scattering members have more significant evolution of their orbits, like that of 2006 SQ$_{372}$ ($q = 24.2$~au, $a = 796$~au; \citealt{Kaib:2009}), while still spending some fraction of their orbit in the inner Oort region.  

Such large-$a$ scattering orbits as FE$_{72}$ provide a conceptual link between the scattering disk and the inner Oort cloud, both past and present.
The emplacement of the scattering disk and its subsequent decay under encounters with Neptune require many millions of minor planets to have been placed on exceptionally large-$a$ orbits \citep{gladman05, levison06}.

The combination of the existence of diffusion in so many of the large-$a$, high-$q$ TNOs and the way in which the scattering disk overlaps with the inner Oort cloud led \citet{bannister17} to the proposal of a mechanism for populating this region, which follows entirely from known physics and the existence of the known planets:
\begin{quote}
An object scatters outward in the initial emplacement of the scattering disk, pushing the orbital semimajor axis into the inner fringe of the Oort cloud. At a semimajor axis of a thousand or more au, Galactic tides couple and torque out the orbit's perihelion. Once an object is orbiting with $q = 50$~au and $a\sim1000-2000$~au, it diffuses to a lower-$a$ orbit via planetary energy kicks. A reservoir population of objects must then exist that cycles under diffusion with $q = 40-50$~au and $a\sim1000-2500$~au.
\end{quote}
The scattering-to-diffusion scenario made a prediction for future large-$a$ discoveries: 
\begin{quote}
    Our scenario for forming 2013 SY99's orbit does show that for an inner Oort cloud object with $q$ lifted to $\gtrsim 55$~au, diffusion will be too weak to retract the semimajor axis. Thus, future discoveries with $q\sim60$~au should have $a \gtrsim 1000$~au.
\end{quote}
The next year, \citet{sheppard18} reported the discovery of the first TNO with perihelion intermediate between SY$_{99}$ and the two very high-$q$ TNOs (Sedna and VP$_{113}$): 2015 TG$_{387}$ has $q=65 \pm 1$~au. 
This TNO has $a = 1190 \pm 70$~au \citep{sheppard18}, in line with the scattering-to-diffusion scenario.
Under the scenario outlined above, its orbit is fossilized. \citet{sheppard18} show that in the  configuration of the known Solar System, TG$_{387}$'s orbit is presently stable, with Galactic tides cycling $q$ on very long (Gyr) timescales.
Potentially, orbits like TG$_{387}$'s could return to the more actively diffusing part of the proposed cycling population.
If stellar flybys are also modelled, TG$_{387}$'s orbit can have its perihelion driven to lower values of $q \sim 50-55$~au by the combination of tidal and stellar perturbations. In this case, TG$_{387}$'s orbit becomes more actively altering, diffusing in $a$ on order of a hundred au or more \citep{sheppard18}. 

The scattering-to-diffusion scenario has an inherent limit to the most distant perihelion orbit it can explain: the kicks from Neptune that permit diffusion to a lower semimajor axis eventually become too weak. 
Thus, diffusion does not explain the $q \sim 80$~au orbits of VP$_{113}$ and Sedna. 
However, it provides an interesting possibility that explains well the orbits of the remainder of the currently known large-$a$ high-$q$ TNOs in the Solar System as we know it.

\section{Dynamical effects expected to be imprinted on the distant Kuiper belt by the presence of an additional massive planet}

In this section, we discuss the results of published n-body simulations that take into account the strong pericentre-raising effects of an additional distant planet \citep[e.g.][]{lawler2017}.
The presence or absence of a massive, distant planet results in very different orbital distributions for large-$a$ TNOs.  
It was originally proposed by \citet{trujillosheppard14} that an apparent clustering in $\omega$ for the six known high-$q$ TNOs at the time could be explained by an undiscovered planet in the distant Solar System.
This theory was expanded on by \citet{batyginbrown16}, who proposed that certain orbits for this distant planet will cause large-$a$, high-$q$ TNOs to have their orbits physically aligned, so the longitude of the ascending node $\omega$, the argument of pericentre $\Omega$, and the longitude of pericentre $\varpi$ (where $\varpi=\Omega+\omega$) will remain confined for all time.
As more high-$q$ TNOs have been discovered, the statistical strength of clustering in all of these orbital angles has grown weaker; through modifications to which orbital $a/q$ cuts are applied, some continue to argue \citep[e.g.][]{brown19} that a clustering signal remains in one or more orbital angles.
Other simulations have highlighted dynamical effects that a massive distant planet would have on this detached TNO population that were not highlighted in the initial published theories \citep[e.g.][]{shankman17}. 

\citet{lawler2017} used n-body simulations to create a Kuiper belt analogue in the presence of a distant massive planet and the four known giant planets, focusing on realistically creating the scattering TNOs using the method of \citet{kaib11b}, including Galactic tides and stellar flybys.  
These simulations also demonstrated that a distant massive planet will take an initially dynamically cold distribution of TNOs and raise pericentres and inclinations on Gyr timescales while creating the scattering disk.
The resulting TNO distributions from these 5-planet emplacement simulations were then compared with a control simulation that included just the known planets \citep{kaib11b}, which have been shown to reproduce the orbital properties of the scattering TNOs at all $a$ \citep{shankman13,lawler2018scat}. 
The 5-planet simulations easily produce a large population of high-$q$ TNOs, but simultaneously produce a wide distribution of inclinations, including a large fraction of retrograde scattering and detached TNOs.  Although substantive in size, \citet{lawler2017} conclude that such orbits would not be strongly detectable in current surveys.

\citet{shankman17} showed that the same inclination-raising mechanism will cause all (then) known high-$q$ TNOs to flip to retrograde inclinations on Gyr timescales, thus there should be a nearly equal number of retrograde as prograde high-$q$ TNOs.
They also showed that with such a broad inclination distribution, the detection of just one of these objects, Sedna, requires a massive number of TNOs on similar $a$ and $q$ orbits spread over a range of inclination, implying a total mass of order tens of Earth masses on such obits. 
\citet{batygin19} also discuss this effect and find it provides a reasonable explanation for known highly inclined TNOs but do not provide detectable population estimates as these would be highly dependent on particulars of the model and not well constrained by current observations.
\citet{Lietal2018} showed inclination flipping will continue to occur for a moderate-eccentricity ($e \gtrsim 0.4$) and near-coplanar distant planet, in a similar mechanism to the near-coplanar flip induced in a hierarchical three-body system \citep{Li:2014}.
Dynamical simulations of the newly discovered high-$q$ TNO described in \citet{sheppard18}, 2015~TG$_{387}$, agree with the existence of inclination-flipping; a large fraction of clones of 2015~TG$_{387}$ in simulations that include a distant giant planet flip to retrograde orbits on Gyr timescales.

While observational constraints on a large retrograde population are currently weak due to their large predicted distances \citep{lawler2017,lawler2018}, these simulations imply that if there is a giant distant planet, the inclination distribution of high-$q$, large-$a$ TNOs should be nearly isotropic (though \citealt{Lietal2018} find some substructure will occur for $a \lesssim 300$~au).
As yet, there remains little evidence of such a dynamically hot inclination distribution.
The highest-$q$ known TNOs both have $i<25^{\circ}$. 
Promisingly, the highest-$i$ TNO yet known, 2015 BP$_{519}$ has $i=54^{\circ}$; however, it is on an orbit actively interacting with Neptune ($q = 35.25 \pm 0.08$~au, $ a= 449.0 \pm 0.5$~au; \citealt{becker18}).
Additionally, the masses required for detection of even one high-$q$ isotropic TNO are worryingly high \citep{shankman17}. 

An additional giant planet is one possible way to explain the orbits of high-$q$ TNOs, but some of the other effects it would have on the orbits of TNOs do not appear to agree with observations.
The science driver behind this latest cycle of additional giant planet simulations was initially proposed was to explain the apparent simultaneous clustering of the three orbital angles ($\Omega$, $\omega$, $\varpi$) of these high-$q$ TNOs, and here we must discuss the complicated and unintuitive biases that are introduced by surveys of this observationally challenging TNO population.

\section{Detectability of orbital effects}

All observational surveys contain biases.
By understanding and carefully keeping track of as many biases as possible, one can understand which types of detections (in this case, which types of orbits) were most unlikely in a survey, and thus which classes of objects represent larger populations than a survey's raw number of detections naively suggest.
Accounting for the fraction of time that a given TNO is visible on its orbit and the survey's sky coverage are the biggest effects, and attempts have been made to quantify and account for biases in several TNO surveys at this level
\citep[e.g.][]{schwambetal10,adams14}.

The OSSOS Ensemble of surveys \citep{petit11,alexandersen16,petit17,bannister18} was specifically designed with bias characterization as a top priority, resulting in Survey Simulator software \citep{petit2018} that allows TNO orbital distribution models to be forward-biased by all the characteristics of the survey, including sky pointing for each survey block, magnitude limits, detection efficiencies and chip gaps.
The OSSOS Ensemble of surveys also took great pains to track every single TNO that was detected, using careful orbital measurements over 5 months in each discovery year and recovery over $>3$ oppositions \citep{bannister18}, so there is no bias in orbit type, unlike other surveys which preferentially do not track low-$q$ TNOs, or have a high rate of lost TNOs.
Because of this, TNOs that were detected as part of the OSSOS Ensemble can be analyzed statistically, and a degree of de-biasing can be achieved for each subpopulation, measuring orbital properties and size distributions \citep{lawler2018}.  The information needed to perform analysis using the OSSOS Ensemble is freely available \citep[see][]{bannister18} and given that this sample represents about 40\% of the currently known TNOs with reliable orbits, the reader is encouraged to consider this particular sample in examinations of the TNO orbital structure.

\subsection{Biases in the angle of pericentre detection in the large-q large-a TNO sample}

\begin{figure*}[htbp]
\centering
\includegraphics[width=\textwidth]{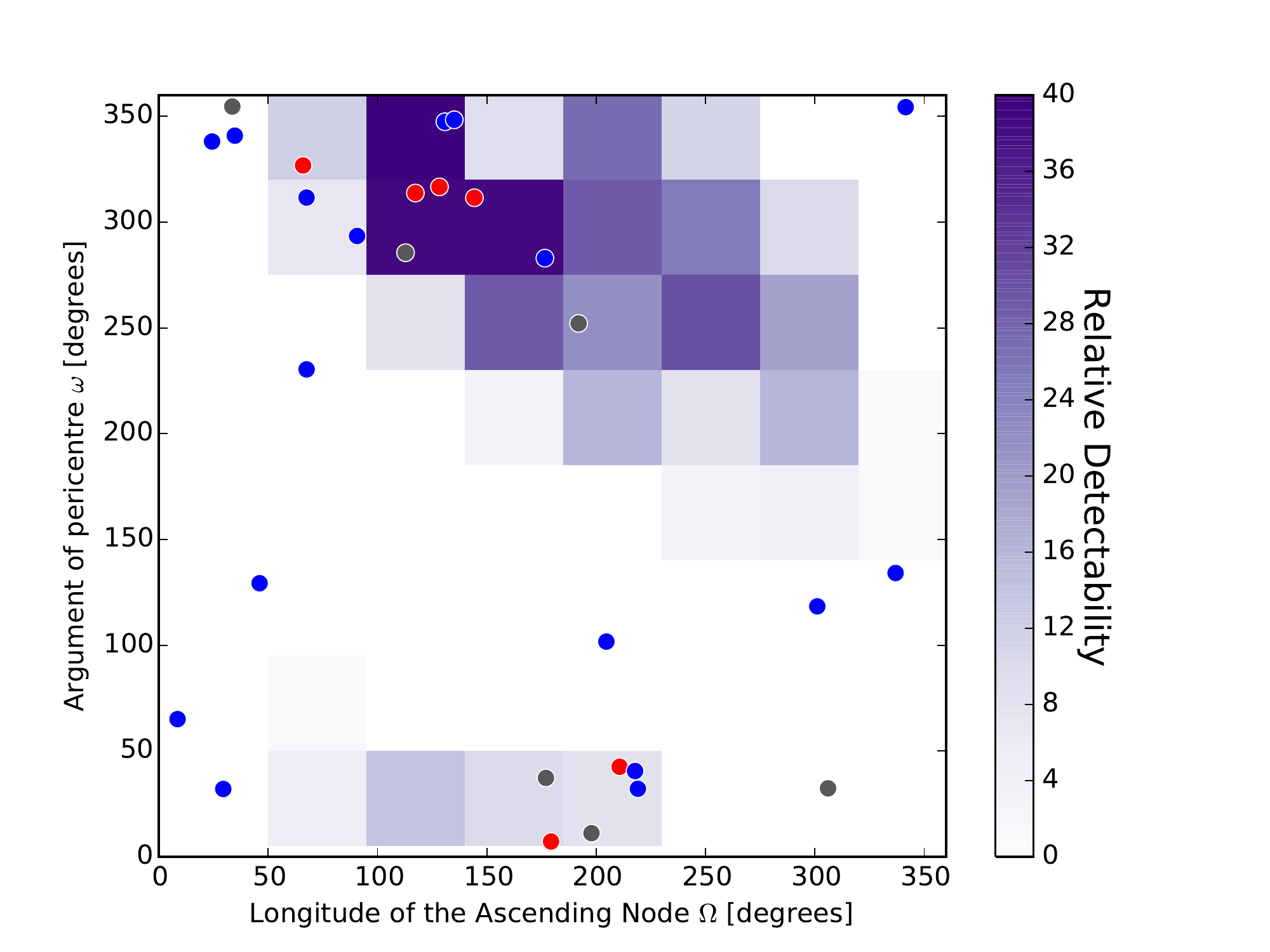}
\caption{Dots show distribution of known TNOs (observed arcs longer than 10 months) with $a > 150$~au and $q > 30$~au in ($\Omega$, $\omega$).  Red dots indicate the first six such known TNOs, grey dots the next six, and the blue dots the most recently discovered TNOs.  The lack of objects with values of $\omega$ near $180^{\circ}$ is not an easy bias to disentangle.  The grid of coloured boxes indicates the number of objects a mock survey would have detected on a grid of $\Omega$, $\omega$ values, sampled from the measured $a$, $q$, and $i$ elements of known TNOs, simulating a southern hemisphere survey.
The most detectable area of ($\Omega$, $\omega$)-space (darkest purple squares) occurs where most of the first known high-$q$ TNOs are. }
\label{fig:bias_jj}
\end{figure*}

Much has been made of the alignment of the pericentre angles of large-$a$ TNOs.  
When all of the high-$q$ TNOs detected in the OSSOS Ensemble are analyzed separately from all other known high-$q$ TNOs, the distribution of orbital angles $\omega$, $\Omega$, and $\varpi$ are consistent with a uniform distribution \citep{shankman17bias,bannister18} 
\citet{brown19} examine the OSSOS Ensemble along, using the characterization information provided with the sample, and also find that, while the sample is not inconsistent with alignment, the sample does not require to be drawn from an aligned distribution.
To restate this a different way, when a uniform distribution of high-$q$, large-$a$ orbits is forward-biased by the OSSOS survey pointings and detection efficiencies, it produces sets of high-$q$, large-$a$ simulated detections that are statistically indistinguishable from the real survey detections.
The high-$q$, large-$a$ TNOs detected by the OSSOS ensemble of surveys show no evidence for orbital clustering in any of the three orbital angles ($\Omega$, $\omega$, $\varpi$).

Many of the high-$q$, large-$a$ TNOs discovered to date are from surveys that have not yet reported their pointing history or tracking fraction, so one cannot statistically test the populations in the same way as the OSSOS detections, but we can make some assumptions about telescope pointing in order to test the biases that are likely present in some of these surveys.
In Figure~\ref{fig:bias_jj} we present the current sample of such
orbits (as of 2018-Oct-1) to allow some examination of that sample.
The figure presents the sample of 6 high-$q$ TNO orbits that created the original
speculation (red points), the next six TNOs detected (grey
points), and the most recently discovered TNOs in blue.  The feature
that originally drew the attention of \citet{trujillosheppard14} was the detection of
objects with $\omega$ near 0$^\circ$ and a complete lack of detections with $\omega$ near 180$^\circ$.
Flux bias in the detected sample
causes most detections to be of TNOs near the pericentres of their
orbits (i.e.\ with mean anomaly $\mathcal{M}$ near $0^{\circ}=360^{\circ}$).  This is coupled with the
habit of conducting TNO searches in fields that predominantly
straddle the sky location of the ecliptic plane, which forces most
discovered TNO orbits to have $\omega$ near $0^{\circ}$ and $180^{\circ}$ (as described in Section~\ref{sec:intro}).
Although the preference for angles near $\omega=0^{\circ}$ is clearly present in the
early sample, there were no detections found near $\omega=180^{\circ}$, which is a puzzling feature of the sample.

In Figure~\ref{fig:bias_jj} we also give, on a grid of ($\omega$, $\Omega$) values, the relative
number of detections one might expect at the given $\Omega$ and
$\omega$ values when drawing from a sample that is uniformly
distributed but with $a$, $q$, and $i$ sampled from the known TNOs.
For simplicity, we assume a flux-limited survey focusing on fields south of the
ecliptic and observing in September, October, November and February
and March (when the best weather conditions occur in the mountains of Chile).  
From the grid of numbers we can see that there are parts
of the ($\Omega$, $\omega$) space where these orbits are much more strongly detectable than others.  

This hypothetical survey is examined as a thought experiment to alert the reader to the complexity of the bias interactions. 
The ($\Omega$, $\omega$) alignment first reported is now largely washed out by the increased sample size (see blue points in Figure~\ref{fig:bias_jj}), but there continues to be a paucity of detections near $\omega = 180^{\circ}$.
Without detailed knowledge of the pointing history and careful measuring of a survey's detection and tracking efficiency through the various seasons of observations,
interpretation of Figure~\ref{fig:bias_jj} is problematic at best.  Regardless of the
distribution's physical reality, there are as yet no described dynamical
processes that keep $\omega$ values away from $180^{\circ}$, and
accepting that the $\omega$ distribution is most likely due to observational biases is
the only supportable explanation.

Subsequent to the claim of an alignment of $\omega$ values, possible
alignment in the longitude of pericentre ($\varpi=\Omega+\omega$) has become
a popular point of discourse,  the appeal being that one can conceive
of physical processes that might align the values of $\varpi$ \citep[e.g.][]{batyginbrown16},
making this a plausibly physical structure.  However, one must
consider that the observationally biased alignment that exists within the raw detected distribution of $\omega$ values propagates forward into a clustering $\varpi$, as the values of $\Omega$ are not uncorrelated, and $\varpi$ is defined as the sum of the two angles $\Omega$ and $\omega$ (see Figure~\ref{fig:bias_jj}).  Thus, although there
are good proposed physical mechanisms to cause a clustering or alignment of
$\varpi$, the clustering of the observed values of
$\varpi$ is contaminated by the same observational biases discussed
in the previous paragraph.

\section{Summary and Conclusions} 

There are several dynamical effects under active theoretical development to explain the observed high-$q$ TNOs. 
The newest announced high-$q$ TNO, 2015~TG$_{387}$, perfectly falls into the ($a$,$q$) range predicted to be affected by chaotic diffusion as described in \citet{bannister17}.
Among the less-explored dynamical mechanisms, rogue planets appear to create distributions of high-$q$ TNOs that match observations reasonably well \citep{lawler2018,silsbee18}.
Precursor simulations like that of \citet{gladmanchan06} should be revisited in light of
the new high-$q$ TNO discoveries to date.  
These types of simulations produce Sedna-like TNOs without a substantial retrograde TNO population at large $a$.
Raising perihelia to the values of the highest-$q$ TNOs, Sedna and VP$_{113}$, remains a challenge to several of the other proposed mechanisms, though stellar flybys remain a promising route.  
Rogue planet models, however, fail to provide a hole in the peri-centre distribution as appears to be present in the detected sample. Indeed, the authors are not aware of any models that reproduce this feature.

While a distant massive planet is effective
at raising pericentres, it also substantially raises inclinations, and current surveys have not yet reported abundant high-$i$ TNOs. 
The authors of this chapter have already reported some of the problematic orbital evolution effects that an additional massive planet in the outer Solar System would create.  In those works we found that the alignment of orbits caused by a massive external planet are not particularly strong \citep{shankman17bias}  and the signature of such an alignment would be difficult to detect in the current sample of known TNOs \citep{lawler2017}.  
Thus, our expectation is that at present there is {\em not} strong evidence of a massive external perturber.

The lack of TNO detections inside the red box in Figures~\ref{fig:bias} and \ref{fig:rogue}, however, provides an intriguing possibility.  We may be able to exclude the existence of such a planet with present published TNO datasets. There are no TNOs reported with pericentres between 50~au and 75~au and semi-major axis interior to 1000~au.  Indeed, other authors have already remarked on the absence of such orbits \citep[e.g.][]{trujillosheppard14,bannister17}.  Recall that in Figure~1 the grid of coloured boxes provides some measure of the inverse probability of detection of particular orbits, given a survey.  A survey that might have detected a TNO at $a \sim 500$~au and $q \sim75$~au is actually more likely to have detected objects with similar $a$ but smaller values of $q$. The same is true of the other $q>75$~au detections: the lower-$q$ but similar $a$ detections are always more likely.  Thus, the lack of detections in the $50 < q < 75$~au range may be indicating that there really is an absence of TNOs on orbits in this range.  This strongly contradicts models of orbital evolution that include an additional planet, as the gravitational action of such an object would cause TNOs to be distributed across a range of $q$ values at any given moment \citep[Figure~\ref{fig:rogue};][]{shankman17,lawler2017}.  Thus, if the lack of objects in the $50~{\rm au} < q < 70$~au range is real, the hypothesized external planet can be excluded.\footnote{There may be some very specialized orbital configurations of a distant planet that preserve the emptiness of this $q$ zone. As of this writing, none have been proposed.}
In most physical situations, multiple effects are in play at any given point in time. 
Perhaps we should be cautious of requiring reduction to a single mechanism to produce all the complexity of the distant TNO populations across the Solar System's history.

\acknowledgements{The authors thank Brett Gladman (UBC) for useful discussions during the preparation of this manuscript.}

\pagebreak

\bibliography{citations}

\end{document}